\documentclass{aastex62}
\usepackage{natbib}
\usepackage[utf8]{inputenc}
\usepackage{graphicx}
\usepackage{rotating}
\usepackage{amsmath}
\begin{document}

\title{Candidate Wide-separation Companions to Nearby, Dusty Young Stars: {\it Gaia} Weighs In}
\author{Joel H. Kastner}
\altaffiliation{
Rochester Institute of
Technology, 54 Lomb Memorial Drive, Rochester NY 14623 USA
(jhk@cis.rit.edu)}

\maketitle


Certain nearby ($D<120$ pc), dusty young stars appear to be accompanied by lower-mass companions at $\sim$10--45 kau separations \citep{Scholz2005,Torres2006,Kastner2008a,Kastner2011,Kastner2012,Kastner2017}. The primaries of these potential wide binaries are the A-type star HR 4796A and F-type star HD 113766A, both of which host debris disks and are themselves members of visual binaries \citep[separations 7.9$''$ and 1.4$''$, respectively;][and references therein]{Schneider2018,Lisse2017}, and the K-type stars TW Hya, T Cha, and (spectroscopic binary) V4046 Sgr AB. 
The  gas-rich, chemically evolved disks orbiting the last three (K-type) stars 
are widely regarded as primordial in origin \citep[][and references therein]{Sacco2014,Kastner2014}
despite pre-main sequence stellar ages, $\sim$10--20 Myr, 
that far exceed the nominal protoplanetary disk dispersal timescale of a few Myr.  We speculated \citep{Kastner2011,Kastner2012} that the putative distant companions may be related to the apparent longevities and/or rejuvenation of planet-forming disks around these late-type stars \citep[see also][]{Zuckerman2015}. 

The precise parallaxes and proper motions contained in {\it Gaia} Data Release 2 \citep[DR2;][]{Gaia2018} now allow us to establish whether the putative wide-separation companions to all (five) of the aforementioned nearby, young stars are in fact equidistant and comoving with these stars (Table 1). 
For three of these proposed wide pairs --- TW~Hya + 2M1102--34; HR 4796A + 2M1235--39; and V4046~Sgr~AB + GSC~07396 --- the {\it Gaia} DR2 data confirm that the two stars lie at the same distance and are comoving (or nearly so), to within the measurement errors\footnote{The small discrepancies in the proper motions of the A and B components of HR 4796 are perhaps indicative of orbital motion, but the (marginal) discrepancy between the parallaxes of HR 4796A and B is likely spurious. Similar arguments pertain to discrepancies between the proper motions of HD 113766A and B and TYC 8246--2900--1A and B and the apparent discrepancy between the parallaxes of the former pair (Table 1, last four rows).}. We conclude that these three pairs indeed constitute wide-separation binaries with projected separations of 44.3 kau, 13.3 kau, and 12.3 kau, respectively. 

In contrast, the DR2 data reveal that 2M1155--79, the proposed distant, comoving companion to T Cha \citep{Kastner2012}, lies $\sim$8.5 pc closer to Earth than T Cha, and the two stars have significantly discrepant (though similar) proper motions. We conclude that 2M1155--79 is merely a foreground member of the $\epsilon$ Cha Association, in which T Cha resides
\citep{Murphy2013}. The DR2 data similarly disprove the hypothesis that HD 113766AB and TYC 8246--2900--1 are equidistant and comoving, while revealing that the lower-mass (K-type) TYC 8246--2900--1 has a close, fainter, M-type binary companion. 
As in the case of T Cha and M1155--79, the similar distances and proper motions of HD 113766AB and TYC 8246--2900--1(AB) suggest that they are members of the same moving group, despite discrepant stellar age estimates \citep[$\sim$12 Myr vs.\ $\sim$40 Myr;][]{Zuckerman2015,Kastner2017}.

Future investigations of the three wide young-star pairs confirmed in this work (Table 1) should be aimed at establishing whether they remain bound at the present epoch and whether the presence of long-lived disks orbiting the primaries is linked to the orbital or dissolution timescales of these wide binaries \citep[][]{Kastner2011}.

{\it This research was supported by NASA grants NNX16AB43G (Exoplanets Research Program) and NNX12AH37G (Astrophysics Data Analysis Program) to RIT. }

\newpage


\begin{sidewaystable}
\caption{\sc Candidate Wide-separation Binaries: {\it Gaia} Data Release 2 Results}
\label{tbl:Continuum}

\footnotesize

\vspace{.2in}

\begin{tabular}{cccccccccccc}
\hline
\hline
Star & $\alpha^a$ & $\delta^a$ & $\pi$ & $D$ & $\mu_\alpha$ & $\mu_\delta$ & $G$ & $B_P - R_P$ \\
& \multicolumn{2}{c}{(J2000)} & (mas) & (pc) & (mas yr$^{-1}$) & (mas yr$^{-1}$) & (mag) & (mag)\\
\hline
\hline
\multicolumn{4}{c}{{\it Candidate wide binaries confirmed as equidistant and comoving (or nearly so) in DR2}}\\
\hline
TW Hya (TWA 1) & 11 01 51.819480 (23) & $-$34 42 17.248788 (22) & 16.642$\pm$0.042 & $60.09\substack{+0.15 \\ -0.15}$ & $-$68.389$\pm$0.054 & $-$14.016$\pm$0.059 & 10.43 & 1.66 \\
2M1102$-$34 (TWA 28) & 11 02 09.75074 (140) & $-$34 30 35.78710 (120) & 16.73$\pm$0.21 & $59.77\substack{+0.76 \\ -0.74}$ & $-$68.98$\pm$0.33 & -13.80$\pm$0.28 & 17.71 & 4.69 \\
\hline
HR 4796A (TWA 11A) & 12 36 00.956832 (98) & $-$39 52 10.59067 (100) & 13.91$\pm$0.13 & $71.89\substack{+0.68 \\ -0.67}$ & $-$55.65$\pm$0.18 & $-$23.74$\pm$0.23 & 5.77 & 0.00\\
HR 4796B (TWA 11B) & 12 36 00.469635 (47) & $-$39 52 16.159140 (52) & 14.103$\pm$0.063 & $70.91\substack{+0.32 \\ -0.32}$ & $-$59.236$\pm$0.096 & $-$29.86$\pm$0.12 & 11.96 & 2.28 \\
2M1235$-$39 (TWA 11C) & 12 35 48.863280 (95) & $-$39 50 24.95242 (165) & 14.00$\pm$0.15 & $71.43\substack{+0.77 \\ -0.76}$ & $-$55.94$\pm$0.20 & $-$24.83$\pm$0.41 & 12.96 & 3.05 \\
\hline
V4046 Sgr AB & 18 14 10.486200 (58) & $-$32 47 35.333808 (53) & 13.811$\pm$0.064 & $72.41\substack{+0.34 \\ -0.33}$ & 3.49$\pm$0.11 & $-$52.754$\pm$0.087 & 9.91 & 1.54\\
GSC 07396--00759 & 18 14 22.077672 (52) & $-$32 46 10.949700 (47) & 14.000$\pm$0.052 & $72.43\substack{+0.27 \\ -0.26}$ &  3.08$\pm$0.10 & $-$52.638$\pm$0.080 & 11.92 & 2.17 \\
\hline
\hline
\multicolumn{4}{c}{{\it Candidate wide/hierarchical binaries not equidistant or comoving in DR2}}\\
\hline
T Cha & 11 57 13.291848 (66) & $-$79 21 31.665060 (50) & 9.122$\pm$0.083 & $109.6\substack{+1.0 \\ -1.0}$ & $-$41.99$\pm$0.11 & $-$9.245$\pm$0.080 & 12.97 & 2.31\\
2M1155$-$79 & 11 55 04.599216 (62) & $-$79 19 11.120736 (48) & 9.886$\pm$0.058 & $101.15\substack{+0.60 \\ -0.59}$ & $-$41.18$\pm$0.13 & $-$4.335$\pm$0.086 & 14.81 & 3.36\\
\hline
HD 113766A & 13 06 35.788295 (44) & $-$46 02 02.278535 (39) & 9.003$\pm$0.057 & $111.08\substack{+0.71 \\ -0.70}$ & $-$32.958$\pm$0.087 & $-$17.095$\pm$0.091 & 8.03 & 0.59 \\
HD 113766B & 13 06 35.655552 (48) & $-$46 02 02.108292 (49) & 9.236$\pm$0.063 & $108.27\substack{+0.74 \\ -0.73}$ & $-$35.580$\pm$0.097 & $-$21.279$\pm$0.126 & 8.22  & 0.68 \\
TYC 8246--2900--1A & 13 06 50.218680 (50) & $-$46 09 56.374416 (30) & 10.111$\pm$0.051 & $98.90\substack{+0.50 \\ -0.50}$ & $-$41.891$\pm$0.095 & $-$18.167$\pm$0.066 & 11.60 & 1.48 \\
TYC 8246--2900--1B & 13 06 50.45969 (160) & $-$46 09 56.34540 (130) & 10.14$\pm$0.21 & $98.6\substack{+2.1 \\ -2.0}$ & $-$35.81$\pm$0.33 & $-$20.23$\pm$0.31 & 15.53 &	2.46 \\
\hline
\hline
\end{tabular}

NOTES:\\
a) Numbers in parentheses are {\it Gaia} DR2 positional (RA, dec) uncertainties in microarcseconds. \\

\end{sidewaystable}


\end{document}